\documentclass[twocolumn,pra,showpacs,floats,superscriptaddress]{revtex4}
\usepackage[dvips]{graphicx}
\usepackage{amsmath,bm}
\usepackage{psfrag}

\begin{document}

\title{
Influences
of a longitudinal and tilted vibration
on stability and dewetting of a liquid film}

\author{S.~Shklyaev}
\affiliation{Department of Theoretical Physics, Perm State
University, 15 Bukirev St., Perm 614990, Russia}
\author{A.~A.~Alabuzhev}
\affiliation{Institute of the Continuous Media Mechanics, Ural Branch of the Russian
Academy of Sciences, Perm
614013, Russia} \affiliation{Department of Theoretical Physics, Perm
State University, 15 Bukirev St., Perm 614990, Russia}
\author{M.~Khenner}
\affiliation{Department of Mathematics, State University of New
York at Buffalo, Buffalo, NY 14260, USA}


\begin{abstract}
We consider the dynamics of a thin liquid film in the attractive
substrate potential and under the action of a longitudinal or a
tilted vibration. Using a multiscale technique we split the film
motion into the oscillatory and the averaged parts. The frequency
of the vibration is assumed high enough for the inertial effects
to become essential for the oscillatory motion. Applying the
lubrication approximation for the averaged motion we obtain the
amplitude equation, which includes contributions from gravity, van
der Waals attraction, surface tension, and the vibration. We show
that the longitudinal vibration leads to destabilization of the
initially planar film. Stable solutions corresponding to the
deflected free surface are possible in this case. Linear analysis
in the case of tilted vibration shows that either stabilization or
destabilization are possible. Stabilization of the dewetting film by 
mechanical action (i.e., the vibration) was first reported by us in 
PRE {\bf 77}, 036320 (2008). This effect may be important for applications.
\end{abstract}



\date{\today}

\maketitle

\section{Introduction}

Dynamics of thin liquid films was extensively studied
during the last decade both experimentally
and theoretically.
The importance of such studies is emphasized by the needs of
modern nano and microfluidic technologies, which commonly employ
films in the $100-1000 \AA$ thickness range. Reviews focusing on
different subfields of research include Refs.
\cite{SHJ,Oron,Eggers}, as well as twelve reviews focusing on
wetting in the recent volume [Annu. Rev. Mater. Res. {\bf 38},
(2008)].

It is well-known that very thin liquid films tend to dewet from
the substrate (rupture). The primary cause for dewetting is the
attractive van der Waals interaction of the film and the
substrate. Loss of stability and rupture of liquid sheets is often
undesirable and may lead to a technological or manufacturing
process failure. Thus understanding dewetting and finding means to
control it are the important and challenging problems.

One of the frequently used methods for controlling the fluid flow
on small-to-large spatial scales is the application of the high frequency
vibration \cite{AverBook,TVC,LChbook}. Several phenomena may emerge when such vibration is
applied, such as the
oscillatory (pulsatile) fluid motion (Faraday instability) and the
time-averaged fluid motion.
Analyses of the pulsatile motion of the liquid layer and thin drops
are carried out in Refs.~\cite{LChbook,wolf-70,Mancebo} for the
transversal vibration and in
Refs.~\cite{LChbook,Hocking_Davis,Oron_Gottlieb} for the longitudinal
one.


The standard high-frequency approximation is based upon the assumption
of the vibration frequency so large, that the viscosity is
important only in a thin boundary layer near the rigid wall
\cite{AverBook,TVC}. This approximation works well for macroscopic
films \cite{LChbook,Lapuerta,Thiele}, but it fails for the thin
films. Nonetheless, we have recently demonstrated \cite{PRE_vert}
that a hierarchy of typical times allows for the averaged
description in a thin film system. Instead of using the standard
high-frequency approximation, we assume that the vibration period
$2\pi/\omega$ is (i) of the order of the characteristic time of
the transversal transfer of the momentum, $\hat H_0^2/\nu$, and
(ii) is small compared to the typical ``horizontal time'',
$L^2/\nu$. (Here $\hat H_0$ is the mean film thickness, $\nu$ is
the kinematic viscosity, and $L$ is the typical horizontal scale.
$L\gg H$ for the thin film.)

Reference~\cite{PRE_vert} develops the averaged description for
the case of the vertical vibration of the substrate. We show that
the influence of the vibration is finite if the amplitude is large
in comparison with $\hat H_0$. In this case the vibration is the
efficient way to stabilize the film against the van der Waals
rupture. In this paper the approach of Ref. \cite{PRE_vert} is
extended to a longitudinal and a tilted vibration.

The paper is organized as follows. In Sec.~\ref{sec:first} the
problem is formulated: the governing equations and the
dimensionless parameters are introduced for the case of the
longitudinal vibration. Also in this section, using the separation
of the time scales, we split the nonlinear boundary value problem
for the fluid flow into two coupled boundary value problems for
the pulsatile and for the averaged flows. The pulsatile flow is
analyzed in Sec.~\ref{sec:puls}. The averaged amplitude equation
for the film height is obtained in Sec.~\ref{sec:aver} using the
solution of the pulsatile flow. The linear stability problem for
the amplitude equation, the weakly nonlinear analysis and the
numerical results on film dynamics are presented in
Sec.~\ref{sec:dynamics}. Stability of the pulsatile flow is
demonstrated in Section~\ref{sec:stab_puls}. This stability
translates into a stability of the averaged flow and thus the
averaged amplitude equation is validated. In Sec.~\ref{sec:tilted}
the analysis of the previous sections is generalized to the case
of a tilted vibration. The conclusions are presented in
Sec.~\ref{sec:summary}.

\section{Formulation of the problem} \label{sec:first}

\begin{figure}[!t]
\includegraphics[width=6.0cm]{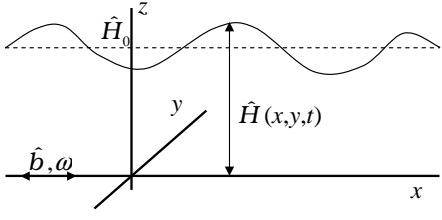}
\caption{Problem geometry: the longitudinal vibration.} \label{fig:1}
\end{figure}

%
We consider a three-dimensional (3D) thin liquid film of the
unperturbed height $\hat H_0$ on a planar, horizontal substrate.
The Cartesian reference frame is chosen such that the $x$ and $y$
axes are in the substrate plane and the $z$ axis is normal to the
substrate (Fig.~\ref{fig:1}).

%
%
The substrate is subjected to the longitudinal harmonic vibration
of the amplitude $\hat b$ and the frequency $\omega$. We assume
that the system is not confined in the horizontal directions
or the vertical boundaries are motionless. Thus the substrate
motion induces the fluid motion due to viscosity and sloshing
modes are not excited.

We assume the film height sufficiently small, so that the
intermolecular interaction becomes important. In this paper, as in
the preceding paper \cite{PRE_vert} we consider the van der Waals
attractive potential. Generalization to other models of wetting
interactions is straightforward. Using same scalings as in
Ref.~\cite{PRE_vert} [i.e., the units for the time, the length,
the velocity and the pressure are $\hat H_0^2/\nu, \ \hat H_0, \
\nu/\hat H_0, \ \rho (\nu/\hat H_0)^2$, respectively; $\nu$ is the
kinematic viscosity and $\rho$ is the density of the liquid] we
begin with the following dimensionless boundary-value problem:
\begin{subequations}
\label{base_eq}
\begin{eqnarray}
\nabla \cdot {\bf v} &=& 0,
\\
{\bf v}_t + {\bf v}\cdot \nabla {\bf v} &=& - {\bf
\nabla} p + \nabla ^2 {\bf v} -G_0 {\bf e_z},
\end{eqnarray}
\end{subequations}
\vspace{-8mm}
\begin{subequations}
\label{base_bcs}
\begin{eqnarray}
\label{bcs_velo}{\bf v} &=& B\Omega\sin\Omega t {\bf e}_x \ {\rm at} \ z = 0, \\
\nonumber h_t + {\bf v} \cdot {\bf \nabla} h &=& w,\\
 \left(p+\phi - Ca
K\right){\bf n} &=&  {\bf n}\cdot {\bf T}  \ {\rm at} \ z =
h(x,y,t).
\end{eqnarray}
\end{subequations}
Here, ${\bf v}=({\bf u},w)$ is the fluid velocity (where ${\bf u}$
is velocity in the substrate plane and $w$ is the component normal
to the substrate), $p$ is the pressure in the liquid, $\bf T$ is
the viscous stress tensor, $h$ is the dimensionless height of the
film, ${\bf e}_{x,z}$ are the unit vectors directed along the $x$
and $z$ axes, respectively, ${\bf n}= \left({\bf e}_z-\nabla
h\right)/\sqrt{1+\left(\nabla h\right)^2}$ is the normal unit
vector to the free surface, $K=\nabla \cdot {\bf n}$ is the mean
curvature of the free surface, $\phi=-A/h^3$ [where
$A=A'/(6\pi\rho\nu^2\hat H_0)$ is the non-dimensional Hamaker
constant], $Ca=\sigma \hat H_0/(\rho \nu^2)$ is the capillary
number (where $\sigma$ is the surface tension), $G_0=g_0 \hat
H_0^3/\nu^2$ is the Galileo number, $B=\hat b/\hat H_0$ is the
non-dimensional amplitude, and $\Omega=\omega \hat H_0^2/\nu$ is
the non-dimensional frequency.

We consider the nonlinear evolution of the large-scale
perturbations. Proceeding exactly as in Ref.~\cite{PRE_vert}, we
first introduce a small parameter $\epsilon$, which is of the order
of the ratio of the mean height $\hat H_0$ to the perturbation
wavelength, i.e. $\epsilon \ll 1$ for long waves. Next, we
introduce conventional stretched coordinates and the time
$$
X=\epsilon x, \ Y=\epsilon y, \ T=\epsilon^2 t, \
$$
assume large capillary number $Ca=C\epsilon^{-2}$, and then
separate the pulsations depending on the ``fast'' time $\tau\equiv
\Omega t$  and the averaged variables which
depend on the ``slow'' time $T$. The detailed analysis of
this procedure is presented in \cite{PRE_vert}; it results in
\begin{subequations} \label{rescaling}
\begin{eqnarray}
\label{rescaled_velo} {\bf u}&=&\epsilon \bar {\bf U} + \tilde
{\bf U}, \ w=\epsilon^2 \bar W + \epsilon \tilde W,
\\
\label{pres_scal} p &=& \bar p + \epsilon^{-1}\tilde p, \ h = \bar
h + \epsilon \tilde h,
\end{eqnarray}
\end{subequations}
where
all fields are $O(1)$
quantities. The pulsations (averaged variables) are marked by tildes (overbars).
Substitution of Eqs.~(\ref{rescaling}) in Eqs.
(\ref{base_eq}) and (\ref{base_bcs}) gives two sets of equations
and boundary conditions for the
pulsational and the averaged parts of the velocity, pressure and height.\\
(i) For the pulsations:
\begin{subequations}
\label{puls_exp}
\begin{eqnarray}
\tilde W_Z &=& -{\bf \nabla}\cdot \tilde {\bf U},\ \Omega \tilde
{\bf U}_{\tau} = - {\bf \nabla} \tilde p + \tilde {\bf U}_{ZZ} ,
\\
\tilde p_Z &=&  0, \label{puls_p_z}
\\
\tilde {\bf U} &=& B\Omega\sin\tau {\bf e}_x, \ \tilde W = 0 \ {\rm at} \ Z = 0, \\
\nonumber \Omega \tilde h_{\tau} &=& - \tilde {\bf U} \cdot {\bf
\nabla}\bar h + \tilde W, \\
 \tilde {\bf U}_Z &=&0, \ \tilde p = 0
\ {\rm at} \ Z = \bar h,
\end{eqnarray}
\end{subequations}
Hereafter ${\bf \nabla}\equiv(\partial_X,\partial_Y,0)$ is a
two-dimensional projection of the gradient operator onto the $X-Y$
plane.

(ii) For the averaged parts:
\begin{subequations}
\label{aver_exp}
\begin{eqnarray}
\bar W_Z &=& -{\bf \nabla} \cdot \bar {\bf U}, \  \bar p_Z = -G_0,
\\
\bar {\bf U}_{ZZ} &=& {\bf \nabla} \bar p + \langle \tilde {\bf U}
\cdot {\bf \nabla} \tilde {\bf U} + \tilde W \tilde {\bf U}_Z
\rangle,
\\
\bar {\bf U} &=& \bar W = 0 \ {\rm at} \ Z = 0, \\
\nonumber \bar p &=&- \langle\tilde p_Z \tilde h\rangle
-\phi(\bar h) - C \bar \nabla^2 h, \\
\nonumber \bar h_T &=& - \bar {\bf U} \cdot {\nabla} \bar h -
\langle \tilde {\bf U} {\bf \nabla} \tilde h \rangle   + \bar W +
\langle \tilde W_Z \tilde h \rangle,\\
\bar {\bf U}_Z &=&-\langle \tilde {\bf U}_{ZZ} \tilde h\rangle
 \ {\rm at} \ Z = \bar h. \label{aver_dyn}
\end{eqnarray}
\end{subequations}
In the set (\ref{aver_exp}) the angular brackets denote averaging
with respect to the fast time $\tau$. The boundary conditions at the
free surface have been shifted at the mean position $\bar h$.
Moreover, we neglect all terms of order $\epsilon$ as they are
unimportant for the further analysis.
Note that the boundary value problem governing the oscillatory
motion, Eqs.~(\ref{puls_exp}) is linear despite the finite
intensity of the oscillatory motion, see
Eq.~(\ref{rescaled_velo}). Also it can be seen that the set
(\ref{puls_exp}) is decoupled from the set (\ref{aver_exp}) and
thus the solution of the former set can be immediately found. It
is worth noting that $B=O(1)$ (in contrast to
Ref.~\cite{PRE_vert}, where the amplitude of the vibration has to
be large in order to provide a finite intensity of the
longitudinal motion.)

\section{Pulsatile motion}\label{sec:puls}
\subsection{Analysis of the general case}\label{ssec:puls}

Here we assume stability of the pulsatile motion (see
Sec.~\ref{sec:stab_puls} for the proof) and determine the solution
of Eqs.~(\ref{puls_exp}). We seek the solution in the form
\begin{subequations} \label{puls_complex}
\begin{eqnarray}
\tilde {\bf U} &=& B \Omega {\bf e}_x{\rm Re} \left[I(X,Y,Z)\exp{\left(i\tau\right)}\right],\\
\tilde W &=& B \Omega {\rm Re} \left[K(X,Y,Z)\exp{\left(i\tau\right)}\right],\\
\tilde p&=&0, \ \tilde h = B {\rm Re} \left[H(X,Y)\exp{\left(i
\tau\right)}\right].
\end{eqnarray}
\end{subequations}
Substitution of this ansatz in Eqs.~(\ref{puls_exp}) gives the set
of equations and boundary conditions governing the amplitudes of
the pulsations:
\begin{subequations}\label{puls}
\begin{eqnarray}
\label{puls_u} K_Z&=&-I_X, \ I_{ZZ}+\alpha^2 I=0, \\
\label{puls_noslip} I&=&-i, \ K=0 \ {\rm at} \ Z=0,\\
i H&=&K - Ih_X, \ I_Z=0 \ {\rm at} \ Z=\bar h,
\end{eqnarray}
\end{subequations}
where $\alpha^2=-i\Omega$.

The solution of the boundary value problem (\ref{puls}) is:
\begin{subequations}\label{sol_puls}
\begin{eqnarray}
\label{sol_puls_I} I&=&-i\frac{\cos\alpha (\bar h-Z)}{\cos\alpha \bar h}, \\
\label{sol_puls_K} K&=&i \bar h_X\frac{1-\cos\alpha
Z}{\cos^2\alpha \bar h},\ H=\frac{\bar h_X}{\cos^2\alpha \bar h}.
\end{eqnarray}
\end{subequations}
Note that the amplitudes $K, I$, and $H$, generally speaking,
depend on $Y$ via $\bar h$, but only the $X$-component of ${\bf
\nabla}\bar h$ is important for the pulsatile motion.

%
\begin{figure}[!t]
\includegraphics[width=8.5cm]{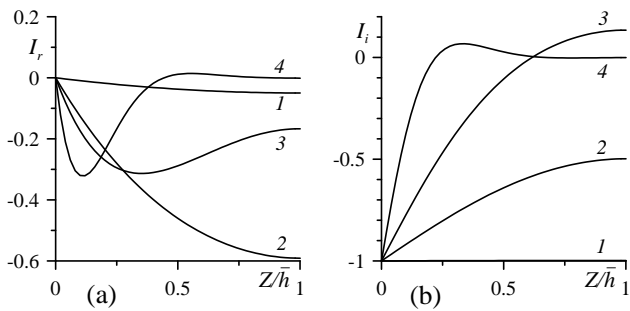}
\caption{Dependence of the real (a) and imaginary (b) parts of
$I$, Eq.~(\ref{sol_puls_I}). Lines 1-4 correspond to $\tilde
\Omega=\Omega \bar h^2=0.1, \, 2,\, 10, \, 100$. } \label{fig:Iz}
\end{figure}
%
\begin{figure}[!t]
\includegraphics[width=6.0cm]{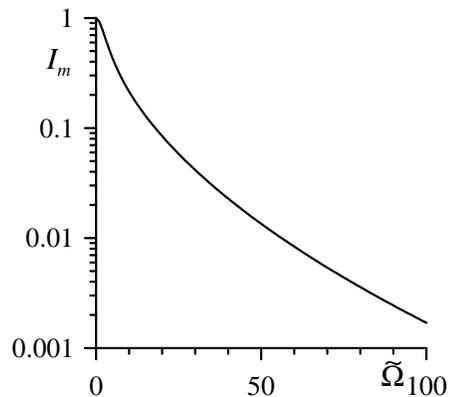}
\caption{The value of $I_m\equiv |I\left(Z=\bar h\right)|$ vs.
$\tilde \Omega$.} \label{fig:Imax}
\end{figure}
%
Figure~\ref{fig:Iz} shows $I_r$ and $I_i$ for various
values of $\Omega$. (Hence and henceforth we use subscripts
``$r$'' and ``$i$'' to denote the real and the imaginary parts,
respectively.) Figure~\ref{fig:Imax} presents $I_m =
|I\left(Z=H\right)|$ at the mean position $\bar h$. Plotting these
figures we use the {\em local} frequency $\tilde \Omega\equiv
\Omega \bar h^2$, which is determined through the local thickness
of the layer. It is obvious that $I_m$ rapidly decays with
increase of the vibration frequency. Note that $I_m^2 \bar h_X$ is
the amplitude of the surface deviation, $|H|$ (where, of course,
$\bar h_X$ is not apriori known).

\subsection{Limiting cases of oscillatory motion} \label{ssec:puls-lim}

{\bf Case $\Omega \gg 1$.} At large $\Omega$ the solution of the pulsatile motion,
Eq.~(\ref{sol_puls}), is small beyond the Stokes boundary layer
adjacent to the substrate. Indeed, taking an obvious relation
$2\cos \alpha \approx \exp(i\alpha)=\exp\alpha^*$ into account
(the asterisk denotes the complex conjugation; here we assume that
$\alpha_r>0$), one immediately arrives at
\begin{eqnarray}
\label{I_high} I&=&  -i e^{-\alpha^* Z}, \\
\label{K_high} K&=& -2i e^{\alpha^*\left(Z-2\bar h\right) }\bar
h_X, \ H=4 e^{-2\alpha^* \bar h} \bar h_X.
\end{eqnarray}
As Eq.~(\ref{K_high}) does not satisfy the no-slip condition, near
the substrate this asymptotic formula has to be rewritten.
Expanding $1-\cos\alpha Z$ in a power series at small
$\alpha Z$ in Eq.~(\ref{sol_puls_K}), we obtain
\begin{equation}
K=2\Omega Z^2 e^{-2\alpha^* \bar h}\bar h_X \ {\rm at} \ Z\to
0.\label{KZ0}
\end{equation}
Of course, Eqs.~(\ref{K_high}) and (\ref{KZ0}) do not match, since
they are the opposite cases ($\sqrt{\Omega} Z \gg 1$ and
$\sqrt{\Omega} Z \ll 1$, respectively) of the high-frequency
approximation, $\Omega\gg 1$, for Eq.~(\ref{sol_puls_K}).

Since only the exponentially small terms are neglected in
Eqs.~(\ref{I_high}),(\ref{K_high}) these asymptotic expressions can be extended
even to moderate $\Omega$ with high accuracy. For instance, the
line 4 in Fig.~\ref{fig:Iz} is indistinguishable from the curve
corresponding to Eq.~(\ref{I_high}a) at $\Omega=100$
($|\alpha|=10$).

{\bf Case $\Omega \ll 1$.} {In the limit of small frequency}
the solution of the
pulsatile problem, Eqs.~(\ref{sol_puls}) reads:
\begin{subequations}\label{sol_puls_low}
\begin{eqnarray}
\label{I_low} I&=&-i-\frac{\Omega}{2}Z\left(2\bar h-Z\right)+ \nonumber \\
&&\frac{i\Omega^2}{24} Z\left(Z^3-4\bar h Z^2+8\bar h^3 \right), \\
K&=&\frac{\Omega}{2}\bar h_XZ^2\left[1+\frac{i\Omega}{12} \left(Z^2-12\bar h^2\right)\right],\\
H&=&\bar h_X\left(1-i\Omega \bar h^2-\frac{2\Omega^2}{3}\bar h^4\right).
\end{eqnarray}
\end{subequations}
Terms up to $\Omega^2$ are held in the expansion of the general
case solution, Eqs.~(\ref{sol_puls}). This accuracy is needed to
provide the averaged effects at low frequency. The expression for
$I$, Eq.~(\ref{I_low}), explains the coincidence of the line 1 in
Fig.~\ref{fig:Iz}(b) with the line $I_i=-1$. Indeed, the
difference between these line is proportional to $\Omega^2\sim
10^{-2}$, which cannot be seen on the scale of the figure.
On the contrary, the real part [see Fig.~\ref{fig:Iz}(a)] is proportional
to $\Omega$ and the corresponding variations are sufficient.

{\bf Case $\Omega$ arbitrary and $\bar h=const.$} In this limit
it is clear from Eqs.~(\ref{sol_puls_K}) that the
oscillatory flow is one-dimensional (1D) and there are no
oscillations of the surface height. Thus in this case the flow is
the oscillatory Couette flow generated by the vibration of the
substrate in a layer with the free surface. Moreover, this flow
differs from the well-known oscillatory Poiseuille flow:
\begin{equation}\label{Poiss}
I=i\left[1-\frac{\cos\alpha (\bar h-Z)}{\cos\alpha \bar h}\right], \\
\end{equation}
only in an additive constant.

It is also important to note that Eqs.~(\ref{I_high}) describe the conventional
``Stokes layer'', i.e. the 1D flow  forced by a high-frequency oscillation of the
rigid plane in a semi-infinite space.

\section{Averaged motion} \label{sec:aver}
\subsection{Analysis of the general case}
Using Eqs.~(\ref{puls_complex}), the problem for
averaged fields, Eqs.~(\ref{aver_exp}),  can be rewritten as
follows (hereafter {\em the overbars are omitted}):
\begin{subequations}\label{aver}
\begin{eqnarray}
\label{hydrostat_aver} p_Z&=&-G_0, \ W_Z=-{\bf \nabla}\cdot {\bf U}, \\
\label{aver_velo} {\bf U}_{ZZ}&=&{\nabla} p +\frac{1}{2} B^2
\Omega^2{\rm Re} \left(I^* I_X+
K^*I_Z\right) {\bf e}_x,\\
{\bf U}&=&W=0 \ {\rm at} \ Z = 0, \\
\nonumber h_T &=&- {\bf U} \cdot {\bf \nabla} h +W -
\frac{1}{2}B^2 \Omega {\rm Re} \left(I^* H\right)_X,\\
\nonumber \ {\bf U}_Z&=&-\frac{1}{2}B^2 \Omega {\rm Re}
\left(I_{ZZ}^*
H\right) {\bf e}_x, \\
p&=&-\phi-C \nabla^2 h  \ {\rm at} \ Z = h. \label{aver_free}
\end{eqnarray}
\end{subequations}
The evolutionary equation for $h$ [the first equation in
Eqs.~(\ref{aver_free})] can be rewritten in the form
\begin{equation} \label{int_aver}
h_T =- \nabla \cdot \int_0^h {\bf U} {\rm d}Z - \frac{1}{2} B^2
\Omega {\rm Re} \left(I^* H\right)_X \ {\rm at} \ Z = h.
\end{equation}

Analytical integration of this set of equations is performed in
Appendix~\ref{app:aver_sol}. It results in the following nonlinear
equation for $h$:
\begin{subequations}\label{h-av_t}
\begin{eqnarray}
h_T&=&{\nabla}\cdot\left(\frac{1}{3}h^3 {\nabla} \Pi_0 \right)-\frac{1}{2}B^2 \Omega^2 \left(Q_1 h^2 h_X\right)_X, \label{hT}\\
\label{Pi_0} \Pi_0 & \equiv&  -\phi(h)-C \nabla^2 h + G_0 h.\\
\label{Q1} Q_1&=&\frac{3\left(2g_1-\gamma g_2\right)}{\gamma^2\left(\cosh\gamma+\cos\gamma\right)^2},\\
\nonumber g_1&\equiv&\sinh\gamma\sin\gamma,\ g_2\equiv
\sinh\gamma\cos\gamma+\sin\gamma\cosh\gamma
\end{eqnarray}
\end{subequations}
where $\gamma=\sqrt{2\Omega}h$. {\it This equation is the central
equation of the paper.} The first three terms at the right-hand
side are the conventional terms resulting from van der Waals
attraction, capillarity and gravity. The term proportional to
$B^2\Omega^2$ is the new term resulting from the longitudinal
vibration of the substrate.

\begin{figure}[!t]
\includegraphics[width=6.0cm]{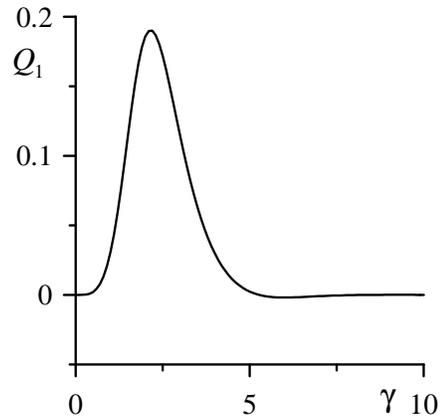}
\caption{Dependence of the coefficient $Q_1$ entering
Eq.~(\ref{h-av_t}) on $\gamma$.} \label{fig:Q}
\end{figure}

The dependence of $Q_1$ is given
in Fig.~\ref{fig:Q}. Note that $Q_1>0$ except for the narrow
interval $5.268<\gamma<8.507$. There is also an infinite set of
such intervals (the second one is at $11.69<\gamma<14.85$), but
the corresponding absolute values of $Q_1$ are very small.

%
%
It is clear that the longitudinal vibration along the $X$ axis
assigns the preferential direction ($X$) in $X-Y$ plane. Thus the
$O(2)$ group of symmetry, which is characteristic for the
amplitude equations governing the thin films dynamics, is broken
for Eqs.~(\ref{h-av_t}). However, the derived amplitude equation
is still invariant under the transformation $X \to -X$.

\subsection{Limiting cases $\Omega \gg 1$ and $\Omega \ll 1$}\label{ssec:limits}

We first notice that the averaged effects vanish in the limiting
case of high frequency, $\Omega \gg 1$. With exponentially small
error, the oscillatory velocity given by Eqs.~(\ref{I_high}) and
(\ref{K_high}) is 1D and uniform along the $X$ axis. Indeed, the
only reason for the variations of the flow velocity stems from the
variation of the film height $h$, and thus this nonuniformity is
exponentially small. As all known mechanisms are based on the
existence of a gradient of the kinetic energy for the pulsations
\cite{TVC}, it is clear that the uniform oscillatory flux is
unable to produce the averaged flow. Therefore, an averaged flow
cannot be produced neither in the boundary layer nor in the core
region. This conclusion agrees well with the asymptotics of $Q_1$
at large $\gamma$:
$$
Q_1\approx -\frac{6}{\gamma} \left(\sin\gamma+\cos\gamma\right)
e^{-\gamma}.
$$

Thus the only relevant limiting case is the case of low-frequency,
$\Omega \ll 1$. Integration of the boundary-value problem for
Eq.~(\ref{aver_velo}) results in the following solution (see
Appendix~\ref{app:aver_sol} for details):
\begin{subequations}\label{dummy}
\begin{eqnarray}
{\bf U}&=&-\frac{1}{2}
Z(2h-Z){\nabla}\Pi_0-\frac{B^2\Omega^2}{2}U^{(v)} h_X {\bf e}_x,\\
U^{(v)}&=&
Z-\frac{\Omega^2Z}{120}\left(Z^4+5hZ^3-20h^2Z^2+80h^4\right).
\end{eqnarray}
\end{subequations}
Then the following equation for the film
height is obtained:
\begin{equation}\label{h-av_low}
h_T={\nabla}\cdot\left(\frac{1}{3}h^3 {\nabla} \Pi
\right)-\frac{1}{15}B^2 \Omega^4 \left( h^6 h_X\right)_X,
\end{equation}
which agrees with the expansion of $Q_1\approx \gamma^4/30$ in
Eq.~(\ref{Q1}) at small $\Omega$.

It is seen in Eq. (\ref{h-av_low}) that at $\Omega\ll 1$ the vibration impact is determined by
the squared amplitude of the pulsatile {\em acceleration}. This
conclusion is expected in view of the similar result for the
vertical vibration.

\section{Film dynamics} \label{sec:dynamics}
\subsection{Linear stability analysis} \label{ssec:stab}

It is convenient to rescale the amplitude equation~(\ref{hT})
using $X=\sqrt{C/3A}\tilde X$, $Y=\sqrt{C/3A}\tilde Y$, and
$T=C/(3A^2)\tilde T$. [Recall that $\phi=-Ah^{-3}$.] This leads
to
\begin{equation}\label{h_t_resc}
h_{\tilde T}={\tilde \nabla}\cdot\left[h^3 {\tilde \nabla}
\left(\tilde G_0 h +\frac{1}{3h^3}- {\tilde \nabla}^2 h
\right)\right]-\tilde V \left(Q_1 h^2 h_{\tilde X}\right)_{\tilde X}, \\
\end{equation}
where $\tilde V=B^2\Omega^2/2A$ and $\tilde G_0=G_0/3A$.

Seeking the solution of Eq.~(\ref{h_t_resc}) in the form $h=1+\xi$,
where $\xi$ is a small perturbation, one obtains:
\begin{equation}\label{xi-av_t}
\xi_{\tilde T}={\tilde \nabla}^2 \left[\left(\tilde
G_0-1\right)\xi - {\tilde \nabla}^2 \xi \right] - \tilde V
Q_1\left(\gamma_0\right) \xi_{\tilde X\tilde X}.
\end{equation}
Here $\gamma_0\equiv \gamma/h=\sqrt{2\Omega}$.

For the normal perturbation $\xi$ proportional to
$\exp\left(i\tilde k_X \tilde X +i \tilde k_Y \tilde Y-\tilde
\lambda \tilde T\right)$ the decay rate $\tilde \lambda$ is:
\begin{equation}\label{decay_resc}
\tilde \lambda=\left( \tilde G_0-1+\tilde k^2\right)\tilde k^2 -
\tilde V Q_1(\gamma_0) \tilde k_X^2,
\end{equation}
where $\tilde k^2=\tilde k_X^2+\tilde k_Y^2$ is the squared
wavenumber.
In terms of the original unscaled variables
$\left(k_X,k_Y\right)$
\begin{equation}\label{decay}
\lambda=\frac{1}{3}\left[ G_0-3A+Ck^2\right]k^2 -\frac{1}{2}B^2
\Omega^2Q_1(\gamma_0) k_X^2.
\end{equation}
%

%
\begin{figure}[!t]
\includegraphics[width=8.5cm]{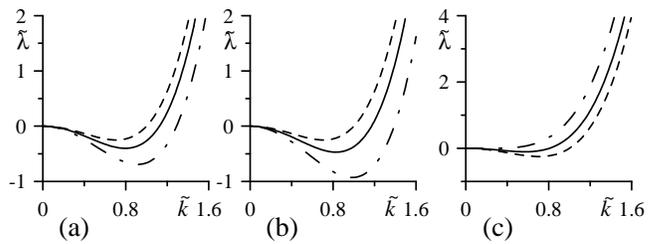}
\caption{Decay rates $\tilde \lambda$ for $\tilde G_0 =3.33\times
10^{-4}, \, \tilde k_Y=0, \, \tilde k_X=\tilde k$. (a)
$\Omega=0.1$. (b)  $\Omega=2$, (c)  $\Omega=20$. (a) and (b):
dashed, solid and dash-dotted lines correspond to $\tilde V =0, \,
200, \, 500$, respectively; (c): dashed, solid and dash-dotted
lines correspond to $\tilde V =0, \, 2, \, 5$, respectively. }
\label{fig:decay}
\end{figure}
%

Since $Q_1$ is positive except for the narrow intervals of
$\gamma$ (see Fig.~\ref{fig:Q}), {\em longitudinal vibration
destabilizes the film} beyond these intervals  [see Fig.~\ref{fig:decay}(a) and
Fig.~\ref{fig:decay}(b)]. Furthermore, one can readily see that
the vibration does not impact the behavior of perturbations with
$\tilde k_X=0$ ({\em longitudinal rolls}). Thus stabilization of
the film by application of the longitudinal vibration is possible only in 2D
systems, where there is no flow in the $Y$-direction, i.e. $\tilde
k=\tilde k_X, \ \tilde k_Y=0$. Below we consider only the behavior of
2D perturbations, which are critical for the reasonable interval
of frequencies.

The typical pictures of the decay rate for this case are shown in
Fig.~\ref{fig:decay}.
For stabilization [Fig.~\ref{fig:decay}(c)] one needs
an extremely large frequency of the vibration. (Recall that
$\Omega=20$ corresponds to $n=300 \, {\rm MHz}$ for the water
layer of the height $1000 \AA$.)

It is obvious that $\tilde k_0$:
\begin{equation}\label{k0}
\tilde k_0^2 =
1+\tilde V Q_1\left(\gamma_0\right)-\tilde G_0
\end{equation}
solves an algebraic equation $\tilde \lambda=0$. For $\tilde k<\tilde k_0$ an
instability takes place.

\subsection{Weakly-nonlinear analysis of 2D systems}\label{ssec:weakly}
Since only the monotonic instability is present, as the analysis
in Sec.~\ref{ssec:stab} confirms, we need to analyze branching of
stationary solutions. Based upon the results of
Sec.~\ref{ssec:stab}, we consider only the 2D system, i.e.
$h=h(\tilde X)$ for the stationary solution. Analyzing the
periodic solutions of a given wavenumber $\tilde k$, we keep in
mind that the obtained results are appropriate for the confined
systems of the length $\tilde L\equiv \pi/\tilde k$.  Indeed, due
to symmetry the boundary conditions $h_{\tilde X}=0$ are imposed
at $\tilde X=0, \tilde L$ for the periodic solution of given
wavenumber $\tilde k$. The same boundary conditions should be set
at the impermeable boundaries $\tilde X=0, \, \tilde L$ for the
confined system. In this case the spectrum of the wavenumbers for
the perturbations is discrete and bounded from below: $\tilde
k^{(n)} \ge \tilde k^{(0)}\equiv \tilde k$, where $\tilde k^{(n)}$
is the $n$th eigen wavenumber corresponding to $(n+1)/2$
wavelength confined into the horizontal length of the system. Thus
for $\tilde k > \tilde k_0$ the longwave instability does not
occur. Moreover, as it is shown below, even for $\tilde k < \tilde
k_0$ the growth of perturbations does not necessarily lead to a
rupture in a confined system.

In the stationary case
Eq.~(\ref{h_t_resc}) can be integrated once. Due to symmetry, the integration constant
is set equal to zero. Thus we obtain
\begin{equation}\label{h_stat}
\tilde k^2 h^{\prime\prime\prime}+\left(h^{-4}+\tilde V \frac{Q_1}
{h}-\tilde G_0\right) h^\prime =0.
\end{equation}
Here the primes denote the derivatives with respect to $\zeta\equiv
\tilde k \tilde X$, i.e. the solution is assumed to be $2\pi$-periodic
in $\zeta$.

To study the weakly nonlinear behavior of the perturbation, we
expand the surface deflection $h$ and the wavenumber $\tilde k$ in
powers of small $\delta$:
\begin{equation}
h=1+\delta\xi_1+\delta^2\xi_2+\ldots, \ \tilde k = \tilde k_0 +
\delta^2 \tilde k_2.
\end{equation}
Substituting these expansions in Eq.~(\ref{h_stat}) we collect the
terms of equal order in $\delta$. The first-order
equation is
\begin{equation}
\hat L \xi_1\equiv \tilde k_0^2
\xi_1^{\prime\prime\prime}+\left[1+\tilde V
Q_1\left(\gamma_0\right) -\tilde G_0\right]   \xi_1^\prime =0.
\end{equation}
Its solution has the form:
\begin{equation}
\xi_1=a\cos \zeta,
\end{equation}
whereas $\tilde k_0$ is given by Eq.~(\ref{k0}).

The second order equation is
\begin{equation}
\label{L1}
\hat L \xi_2=\left(4-\tilde V F_1\right)\xi_1\xi_1^{\prime},
\end{equation}
where
$$
F_1\equiv\gamma_0^2\frac{d}{d\gamma_0}\frac{Q_1\left(\gamma_0\right)}{\gamma_0}.
$$
The solution of Eq.~(\ref{L1}) is
$$
\xi_2=-\frac{4-\tilde V F_1}{12 \tilde k_0^2} a^2 \cos 2\zeta.
$$
The third-order equation is:
\begin{equation}
\hat L \xi_3=\left[\left(4-\tilde V
F_1\right)\xi_1\xi_2-\frac{1}{3}\left(10+\tilde V
F_2\right)\xi_1^3-2\tilde k_0\tilde k_1
\xi_1^{\prime\prime}\right]^{\prime},
\end{equation}
where
$$
F_2\equiv\frac{\gamma_0^3}{2}\frac{d^2}{d\gamma_0^2}\frac{Q_1\left(\gamma_0\right)}{\gamma_0}.
$$
The solvability condition of this equation couples the correction
$\tilde k_2$ and the amplitude of the surface deviation as follows:
\begin{equation}
\tilde k_2=\left[10+\tilde V F_2+\frac{\left(4-\tilde V
F_1\right)^2}{6\tilde k_0^2}\right]\frac{a^2}{8\tilde k_0}.
\end{equation}
If the term in the bracket is positive, then $\tilde
k_2>0$ and the subcritical bifurcation takes place. It is obvious
that the branching solution is unstable in this case. Otherwise, the
supercritical bifurcation occurs and the {\em stable} stationary
solution corresponding to the deflected surface emerges.

%
\begin{figure}[!t]
\includegraphics[width=6.0cm]{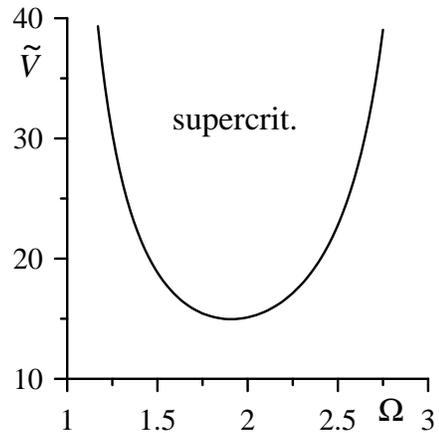}
\caption{The domain in the parameter space, where the supercritical
branching of the surface deviation takes place. $\tilde G_0=0$,
but the curve is almost unchanged even for $\tilde G_0=0.5$.}
\label{fig:sub_super}
\end{figure}
%
The curve separating these two regions in the plane $(\tilde
V,\Omega)$ is shown in Fig.~\ref{fig:sub_super}. It can be readily
seen that the supercritical excitation exists only at the large enough
values of $\tilde V$, when the destabilization effects are well
pronounced. Nevertheless, this phenomenon is quite interesting and
unexpected and requires an additional analysis.

\subsection{Stationary periodic solutions} \label{ssec:stationary}

\begin{figure}[!t]
\includegraphics[width=6.0cm]{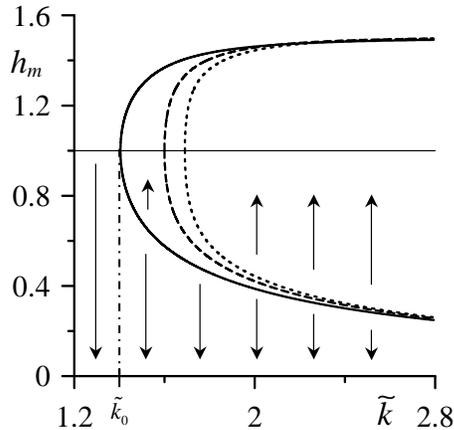}
\caption{Variation of the maximal (the upper branches) and the
minimal (the lower branches) thickness of the layer with $\tilde
k$ for $\tilde V=10$ and $\Omega=1, \, 1.5, \, 2$ -- solid,
dashed, and dotted lines, respectively. For the former case
($\Omega=1$) the value of $\tilde k_0$ is marked. Stability of the
corresponding solutions is indicated by arrows.  }
\label{fig:ampl10}
\end{figure}

\begin{figure}[!t]
\includegraphics[width=8.5cm]{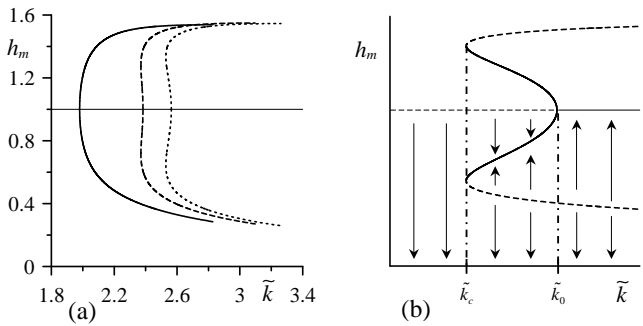}
\caption{Variation of the maximal (the upper branches) and the
minimal (the lower branches) thickness of the layer with $\tilde
k$. (a) -- $\tilde V=30, \ \Omega=1, \, 1.5, \, 2$ -- solid,
dashed, and dotted lines, respectively. (b) -- sketch with marked
$\tilde k_0$ and $\tilde k_c$: solid lines correspond to stable
solutions, dashed lines -- to unstable ones. Domains of attraction
are shown by arrows. } \label{fig:ampl30}
\end{figure}

\begin{figure}[!t]
\includegraphics[width=6.0cm]{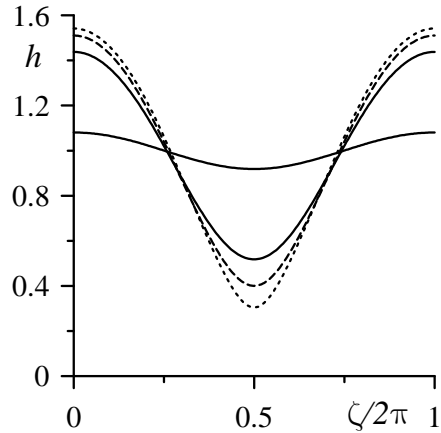}
\caption{The shape of the free surface for $\tilde V=30, \
\Omega=2$; $\tilde k=3, \, 2.7, \, 2.56$ -- dotted, dashed, and
solid lines, respectively. The solid line with smaller
surface deformation corresponds to the stable state, another solid
curve corresponds to unstable solutions as well as dotted and
dashed ones. } \label{fig:hx}
\end{figure}

\begin{figure}[!t]
\includegraphics[width=8.5cm]{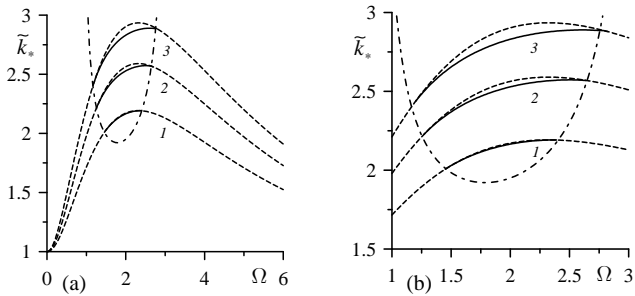}
\caption{Variation of $\tilde k_*$ with $\Omega$ for $\tilde V=20,
\, 30, \, 40$ -- lines 1--3, respectively. Here $\tilde k_*=\tilde
k_0$ correspond to the pitchfork bifurcation (dashed lines) and
$\tilde k_*=\tilde k_c$ (solid lines) correspond to the
saddle-node bifurcations. Dash-dotted lines show the locus of cusp
points, where $k_0=k_c$; these lines correspond to $k_0(V)$ along
the line shown in Fig.~\ref{fig:sub_super}.} \label{fig:kc}
\end{figure}

To study stationary periodic solutions of a finite amplitude we
integrate Eq.~(\ref{h_stat}) with the boundary conditions:
\begin{equation}
h^{\prime}=0 \ {\rm at} \ \zeta=0,2\pi,
\end{equation}
and the mass conservation condition:
\begin{equation}
\int_0^{2\pi}h {\rm d} \zeta=2\pi.
\end{equation}
(Recall that for the equivalent confined system only half of
the period should be taken, i.e. $\zeta<\pi$.) This boundary
value problem was solved by the shooting method. The numerical
results are presented in Figs.~\ref{fig:ampl10}-\ref{fig:kc} for
$\tilde G_0 = 3.33\times 10^{-4}$.

%
%
%
The amplitude curves $h_m(k)$ are shown for $\tilde V=10$ and
$\tilde V=30$ in Figs.~\ref{fig:ampl10} and \ref{fig:ampl30},
respectively. These figures confirm the results of the
weakly-nonlinear analysis: the supercritical bifurcation takes place
for $\tilde V=30$.

In Fig.~\ref{fig:ampl10} the value of $\tilde k_0$ is marked for
$\Omega=1$ (solid line). For smaller $\tilde k$ instability of the
flat surface gives rise to a rupture. The stationary solution
exists only for $\tilde k>\tilde k_0$, i.e. all the curves in
Fig.~\ref{fig:ampl10} represent unstable {\em subcritical}
solutions. The lower branches of the amplitude curves in
Fig.~\ref{fig:ampl10} can be thought of as the boundaries of the
domains of attraction for the equilibrium state ($h=1$) in the
framework of the evolutionary problem, Eq.~(\ref{h_t_resc}) (or
the similar equation with $\zeta=\tilde kX$ being introduced). An
initial distortion of the surface $h_0(\zeta)$ with the trough
deeper than $h_m$ necessarily leads to rupture, while an initial
distortion with $h_0(\zeta)>h_m$ for any $\zeta, \, 0<\zeta <2\pi$
decays with time resulting in the equilibrium state at $T\to
\infty$ (see arrows in Fig.~\ref{fig:ampl10}). Of course, this
interpretation is not exact, as the whole variety of the initial
states $h_0(\zeta)$ is characterized by the only value, $h_m$.
Therefore, the domain of attraction has to be confined by a band of finite thickness.
However, due to the fast growth of the attracting van der Waals
potential with the decrease of $h$, $h_m$ is a perfect
characteristic for the domain of attraction and the band
thickness is rather small. Our numerical tests based on the
finite-difference computation of Eq.~(\ref{h_t_resc}) support this
conclusion.

For larger values of $\tilde V$ the stable distorted surface is
found within some interval, $\tilde k_c<\tilde k<\tilde k_0$, see
the dashed and the dotted curves in Fig.~\ref{fig:ampl30}(a) and a
schematic plot in Fig.~\ref{fig:ampl30}(b). In the latter figure
domains of attraction and stability properties of the obtained
solutions are also demonstrated. Again, the lower branch
$h_m(\tilde k)$ of the unstable solution can be thought of as the
boundary of the domain of attraction: the rupture occurs for
initial distortions with $\min h_0(\zeta)<h_m$; in the opposite
case the initial perturbation decays and the stable branch is
achieved (either the stable branch of solution for $\tilde
k<\tilde k_0$ or $h=1$ for $\tilde k>\tilde k_0$). To the best of
our knowledge, this problem provides the first example where a
\emph{nonplanar free surface} is {\em stable} in the presence of
the {\em attracting van der Waals potential}.

%
%

The interval of existence of a stable solution is quite small and
for $\tilde k=\tilde k_c$ a saddle-node bifurcation takes place,
see Fig.~\ref{fig:ampl30}(b). Both stable (with smaller values of
$|h_m-1|$) and unstable (with larger values of $|h_m-1|$) branches
of the solution disappear at $\tilde k=\tilde k_c$ and do not
exist at smaller $\tilde k$.

The typical shapes of the surface along the dotted curve in
Fig.~\ref{fig:ampl30}(a) is presented in Fig.~\ref{fig:hx}. For
the smaller value of the wavenumber, $\tilde k=2.56$, bistability
takes place, i.e. there are two stationary shapes, the stable and
unstable one (with the smaller and larger surface deformation,
respectively).

The bifurcation lines $\tilde k_0(\Omega)$ and $\tilde
k_c(\Omega)$ are shown in Fig.~\ref{fig:kc}. The stable states
with the deformed free surface exist between the corresponding
solid and dashed lines. The direct pitchfork bifurcation takes
place within this interval of $\Omega$. Otherwise, the branching
is subcritical.

%
%

It should be noted that $\tilde k_c>\tilde k_0^{(c)}$, where
$\tilde k_0^{(c)}\equiv \tilde k_0$ for $\tilde V=0$, i.e. the critical
wavenumber in absence of the vibration. This inequality ensures
that rupture always occurs at $\tilde k<\tilde k_0^{(c)}$,
i.e. there is no stable stationary states even with deformed free
surface. Thus the longitudinal vibration cannot be applied for
``nonlinear stabilization'', i.e. in order to produce stable
distorted film with $\tilde k<\tilde k_0^{(c)}$.

\section{Stability of the time-periodic motion (pulsatile flow)} \label{sec:stab_puls}
\subsection{Reduction of the stability problem}
The finite intensity of the time-periodic solution,
Eqs.~(\ref{rescaling}), (\ref{puls_complex}), and
(\ref{sol_puls}), raises question of the solution stability. In
order to show stability or instability, we return to the unscaled
Eqs.~(\ref{base_eq}) and (\ref{base_bcs}) and represent the
velocity, the pressure and the surface deflection in the form
\begin{equation}\label{pert_puls}
{\bf v}=\tilde {\bf U} + \epsilon \tilde W {\bf e_z}+{\bf V}, \
p=\bar p+P, \ h=\bar h +\epsilon \tilde h + \Upsilon.
\end{equation}
(Within this section we restore the bar for the averaged film
height, $\bar h$.) Here in accordance with Eqs.~(\ref{rescaling})
the dominant parts of the unperturbed velocity are the pulsatile one,
while for the pressure field and the layer height the averaged
parts dominate over the oscillatory ones. Next, we linearize
Eqs.~(\ref{base_eq}) and (\ref{base_bcs}) with respect to the
small perturbations ${\bf V}, \, P$ and $\Upsilon$, assuming 2D
perturbation ${\bf V}=(U,0,W)$. (We will show that such
perturbations are critical in the sense that they occur prior to
the onset of 3D instability.) Thus
we arrive at the following boundary value problem:
\begin{subequations}
\label{pert_puls_eq1}
\begin{eqnarray}
U_x &=&-W_z,
\\
U_t&=&- U_0 U_x -U U_{0x} - W U_{0z}   - P_x + \nabla ^2 U, \\
W_t &=&- U_0 W_x -U W_{0x} - P_z + \nabla ^2 W,\\
U &=& W=0 \ {\rm at} \ z = 0, \\
\nonumber P &=& - \left(\bar p_z + \frac{3A}{\bar
h^4}\right)\Upsilon - Ca K^{(l)} + {\bf T}_{nn}^{(l)},\\
{\bf T}_{nt}^{(l)}&=&0,\ \Upsilon_t + U_0 \Upsilon_x+U \bar h_x =
W  \ {\rm at} \ z = \bar h. \label{dyn_pert1}
\end{eqnarray}
\end{subequations}
Here we set
\begin{equation}
U_0=B\Omega{\rm Re} \left[I(z)\exp\left(i\Omega t\right)\right].
\end{equation}
according to Eq.~(\ref{puls_complex}). Besides, we have neglected
all terms of order $\epsilon$, for example $\tilde W$ and $\tilde
h$. Thus with accuracy $O(\epsilon)$ the base flow can be thought
of as 1D flow.

Components of the viscous stress tensor are linearized as follows
near the unperturbed surface, $z=\bar h(X,T)$:
\begin{eqnarray}
T_{nn}^{(l)}&\equiv& {\bf n}_0 \cdot \left({\bf T}_1 +
\partial_z {\bf T}_0\Upsilon\right) \cdot {\bf n}_0 + 2{\bf n}_0
\cdot
{\bf T}_0 \cdot {\bf n}_1 \\
\nonumber T_{nt}^{(l)}&\equiv& {\bf n}_0 \cdot \left({\bf T}_1 +
\partial_z {\bf T}_0\Upsilon\right) \cdot {\bf t}_0 \\
&&+ {\bf n}_1 \cdot {\bf T}_0 \cdot {\bf t}_0 +{\bf n}_0 \cdot
{\bf T}_0 \cdot {\bf t}_1,
\end{eqnarray}
where ${\bf T}_1$ is calculated over $U$ and $W$, ${\bf T}_0$ is
the viscous stress tensor of the base flow [i.e. $T_{0xx}=2U_{0x},
\ T_{0xz}=U_{0z}, \, T_{0zz}=O(\epsilon)$], ${\bf n}_0$ and ${\bf
n}_1$ (${\bf t}_0$ and ${\bf t}_1$) are the normal (tangential)
vector for the unperturbed free surface $z=\bar h$ and the
correction due to its perturbation $\Upsilon$, respectively:
\begin{eqnarray}
\label{unit_unpert}{\bf n}_0&=&\frac{{\bf e}_z -\bar h_x{\bf
e}_x}{\sqrt{1+\bar h_x^2}},\ {\bf t}_0=\frac{{\bf e}_x +\bar
h_x{\bf e}_z}{\sqrt{1+\bar h_x^2}},
\\
{\bf n}_1&=&-\frac{\Upsilon_x}{\sqrt{1+\bar h_x^2}}\left({\bf
e}_x+\frac{\bar h_x}{1+\bar h_x^2} {\bf e}_z\right),\\
{\bf t}_1&=&\frac{\Upsilon_x}{\sqrt{1+\bar h_x^2}}\left({\bf
e}_z-\frac{\bar h_x}{1+\bar h_x^2} {\bf e}_x\right).
\end{eqnarray}
Finally, for the curvature we obtain:
\begin{equation}
K^{(l)}=-\frac{\Upsilon_{xx}}{\left(1+\bar
h_x^2\right)^{3/2}}+3\frac{\bar h_x \bar h_{xx}
\Upsilon_x}{\left(1+\bar h_x^2\right)^{5/2}}.
\end{equation}

It is clear that the $x$-derivatives of the base state variables
are $O(\epsilon)$  quantities and thus they can be omitted from
the leading-order analysis. Indeed, the base state depends on $x$
only via the slow coordinate $X$, whereas the characteristic
wavelength of perturbations is $O(1)$. For example,
$U_{0x}=\epsilon U_{0X}$. Also, for the same reason the
derivatives with respect to the slow time $T$ are omitted. This
procedure is equivalent to the methods of frozen coefficients:
neglecting $\epsilon$-terms above we assume that the base flow is
1D and {\em locally} $x$-independent. Moreover, the free surface
is almost horizontal: according to Eqs.~(\ref{unit_unpert}) one
obtains ${\bf n}_0={\bf e}_z$ and ${\bf t}_0={\bf e}_x$, which
significantly simplifies the formulas. As a result, the stability
problem for the pulsatile motion reads:

\begin{subequations}
\label{pert_puls_eq}
\begin{eqnarray}
U_x &=&-W_z,
\\
U_t&=&- U_0 U_x - W U_0^{\prime}   - P_x + \nabla ^2 U, \\
W_t &=&- U_0 W_x  - P_z + \nabla ^2 W,\\
U &=& W=0 \ {\rm at} \ z = 0, \\
\nonumber \ P &=& \left(G_0 + \frac{3A}{\bar h^4}\right)\Upsilon -
\frac{C}{\epsilon^2} \Upsilon_{xx}+ 2\left(W_z-U_0^\prime\Upsilon_x\right),\\
\nonumber U_z&=&-W_x-U_0^{\prime} \Upsilon,\\
\label{dyn_puls_pert}\Upsilon_t &=&- U_0 \Upsilon_x + W \ {\rm at}
\ z = \bar h.
\end{eqnarray}
\end{subequations}
%
%
%
The primes denote the $z$-derivatives of $U_0$.
Equation~(\ref{hydrostat_aver}) is taken into account to calculate
$\bar p_z$ entering Eq.~(\ref{dyn_pert1}). Recall that
$C=\epsilon^2 Ca$.

Large factor $\epsilon^{-2}$ in normal stress balance [the first
relation from Eqs.~(\ref{dyn_puls_pert})] results in $\Upsilon=0$,
i.e. the surface is {\em locally undeformable}. This, in turn,
means that the boundary value problem~(\ref{pert_puls_eq}) is the
stability problem for 1D flow, periodic in time. This problem, the
so-called time-dependent Orr-Sommerfeld problem, is analyzed in
Secs.~\ref{ssec:Orr-Sommer} and \ref{ssec:Pois_stab}.

\subsection{Orr-Sommerfeld problem for the pulsatile flow}\label{ssec:Orr-Sommer}

It is well-known \cite{squire-33} that 2D perturbations are
critical for either stationary or time-dependent Orr-Sommerfeld
problem. This proves neglecting the $y$-component of $\bf V$ and
allows to introduce in Eqs.~(\ref{pert_puls_eq}) a streamfunction
$\psi$. Setting
$$
\Upsilon = 0, \ U=-\psi_z, \ W=\psi_x
$$
and separating the $x$-coordinate by means of
$\psi(x,z,t)=\hat\psi(z,t)\exp \left(ikx\right)$ we arrive at
\begin{subequations}
\label{Orr_Zommer}
\begin{eqnarray}
D^2 \hat\psi_t&=&- ik \left(U_0 D^2\hat\psi - U_0^{\prime\prime} \hat\psi\right)+D^4 \hat\psi, \\
\hat\psi &=& \hat\psi^{\prime}=0 \ {\rm at} \ z = 0, \\
\hat\psi&=&D^2\hat\psi=0 \ {\rm at} \ z = \bar h,
\end{eqnarray}
\end{subequations}
where $D^2\hat\psi\equiv\hat\psi^{\prime\prime}-k^2 \hat\psi$. It
is convenient to rescale the vertical coordinate and the time in
such a way that $\bar h=1$. Keeping the same notations for the
rescaled variables we obtain:
\begin{subequations}
\label{Orr_Zommer_resc}
\begin{eqnarray}
D^2 \hat\psi_t&=&- ik \left(U_0 D^2\hat\psi - U_0^{\prime\prime} \hat\psi\right)+D^4 \hat\psi, \\
\hat\psi &=& \hat\psi^{\prime}=0 \ {\rm at} \ z = 0, \\
\hat\psi&=&D^2\hat\psi=0 \ {\rm at} \ z = 1,
\end{eqnarray}
\end{subequations}
where
$$
U_0=R {\rm Re} \left[I_0(z)\exp\left(i\tilde\Omega
t\right)\right], \ I_0=-i\frac{\cos\tilde \alpha (1-z)}{\cos\tilde
\alpha}.
$$
We again use the {\em local} oscillation frequency (cf.
Sec.~\ref{sec:puls})
$$
\tilde\Omega=\Omega\bar h^2, \ \tilde \alpha=\alpha \bar h;
$$
the Reynolds number $R$ is introduced as follows
\begin{equation}\label{Re}
R=B\Omega\bar h.
\end{equation}
First, we show that any constant value $C_0$ can be added to
the amplitude of the velocity oscillations $I_0(z)$. It can be
readily seen that Eqs.~(\ref{Orr_Zommer_resc}) are invariant under
transformation
\begin{equation}\label{transform}
\hat \psi \to \hat \psi \exp\left[\frac{ikR}{\tilde \Omega}{\rm
Re}\left(iC_0 e^{i\tilde \Omega t}\right)\right], \ I_0\to I_0+C_0.
\end{equation}
Because $\hat \psi$ is transformed by the periodical in time
factor, stability properties do not change under this
transformation. In particular, setting $C_0=i$ we reduce $I_0$ to
the velocity profile of the oscillatory Poiseuille flow,
Eq.~(\ref{Poiss}), or
$$
I_0=i\left[1-\frac{\cos\tilde \alpha (1-z)}{\cos\tilde
\alpha}\right]
$$
in terms of rescaled coordinate.

Thus, the stability problem for the pulsatile flow is reduced to
the Orr-Sommerfeld problem (\ref{Orr_Zommer_resc}) for the
oscillatory Poiseuille flow. In spite of the detailed studies of
the modulated Poiseuille flow
\cite{Singer,Straatman,Davis_annu,Kerczek}, there have been no
papers that deal with the particular case of the zero mean value
of the pressure gradient.

Intuitively, the flow in a finite layer is more stable than the
flow in a semi-infinite space, i.e. the Stokes layer, which is known to
be stable \cite{Davis_annu}. Nevertheless, we give some results on
the stability of the flow in Sec.~\ref{ssec:Pois_stab}, which
confirm the above guess.

\subsection{Stability problem for the oscillatory Poiseuille flow}\label{ssec:Pois_stab}
To solve the linear stability problem (\ref{Orr_Zommer_resc}) for the time-dependent
flow, we apply the following method. First, due to Floquet theorem
the amplitude of streamfunction $\hat \psi$ can be represented in
the form:
\begin{equation}
\hat \psi=e^{-\Lambda t} \hat \Psi(z,t).
\end{equation}
Here $\hat \Psi(z,t)$ is the periodic function of time with the
period $2\pi/\tilde \Omega$, which thus can be expanded in a
Fourier series as follows:
\begin{equation}
\hat \Psi(z,t)=\sum_{-\infty}^{+\infty} \Psi_n(z)\exp\left(i\tilde
\Omega n t\right).
\end{equation}
Substituting this ansatz into Eqs.~(\ref{Orr_Zommer_resc}), we
arrive at the chain of coupled boundary value problems for the
Fourier component, $\Psi_n(z)$. Truncation of the series by
replacing the upper and lower limits of summation with $N$($-N$),
respectively, leads to the set of $4(2N+1)$ ordinary differential
equations. This set has been solved by the shooting method. We
followed several lower branches of the spectrum; well-pronounced
stabilization of the flow with increase of $R$ was found for all
the branches.
%
\begin{figure}[!t]
\includegraphics[width=5.5cm]{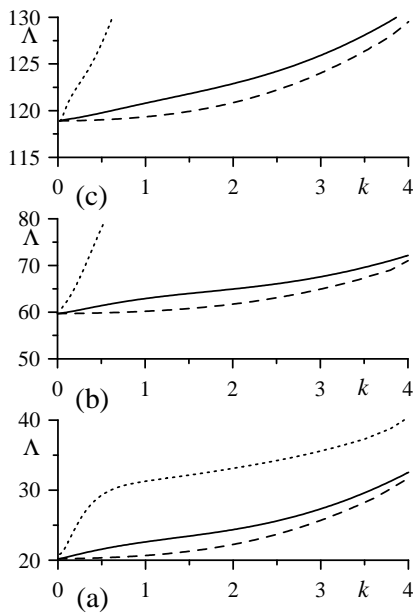}
\caption{Floquet exponents $\Lambda$ vs. $k$ for $Re=50000$. Three
lower branches of the spectrum (a)-(c) for $\tilde \Omega=0.2, \,
1, \,5$ -- dotted, solid, and dashed lines, respectively.}
\label{fig:Pois_stab}
\end{figure}
%

The example of the computations is presented in
Fig.~\ref{fig:Pois_stab}, where the three lower branches of the
Floquet exponent $\Lambda$ are shown. The curves are obtained with
$N=10$; the changes caused by larger $N$ cannot be seen on the
scale of the figure. Therefore, the oscillatory Poiseuille flow is
shown to be stable even for finite values of $\tilde \Omega$.

\section{Vibration along the near-vertical axis} \label{sec:tilted}
\subsection{General notes and the analysis of the pulsatile motion} \label{sec:tilted_first}

\begin{figure}[!t]
\includegraphics[width=6.0cm]{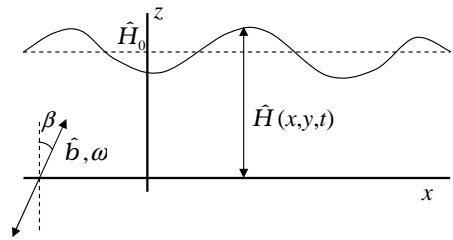}
\caption{Problem geometry: the tilted vibration.}
\label{fig:geom_tilt}
\end{figure}

In this section we briefly generalize the previous analysis
(Secs.~\ref{sec:puls}-\ref{sec:dynamics} and Ref.~\cite{PRE_vert})
to the case of the vibration along the axis that is tilted at a certain angle
$\beta$ to the substrate (see Fig.~\ref{fig:geom_tilt}). It is
obvious that at the finite values of $\beta$ the normal component
of the acceleration is unimportant. Indeed, it is shown in
Ref. \cite{PRE_vert} that the impact of vertical vibration is finite only
at large amplitudes: $B=O\left(\epsilon^{-1}\right)$, while the
horizontal vibration becomes essential even at finite $B$.
Therefore, the longitudinal component of the vibration velocity is
determinative.

%
%
The only case where the effect of the normal acceleration can
compete with the one due to longitudinal motion is the
large-amplitude almost vertical vibration, i.e. $\beta \to 0$.
This case is important for applications because in experiments it
is difficult to ensure the absolutely vertical axis of the oscillatory motion
-- the horizontal components of accelerations occur inevitably.

Thus we assume $\beta$ small, i.e. $\cos \beta\approx 1$
and the near vertical motion of the substrate  according to the law $z=z_0-B\cos\tau$
(in the laboratory
reference frame). The
fluid motion in the reference frame moving vertically with the
substrate is governed by the Eqs.~(\ref{base_eq}) and
(\ref{base_bcs}); the replacements are (i) the gravity modulation:
\begin{equation}
G(t)=G_0+B\Omega^2 \cos \tau
\end{equation}
and (ii) the longitudinal velocity is now $\beta B \Omega \sin
\Omega t {\bf e}_x$ in Eq.~(\ref{bcs_velo}).

Assuming large vibration frequency $B$  we introduce the rescaled
amplitude of the vibration $b\equiv \epsilon^{-1} B$ and the
rescaled angle of the vibration $\mu \equiv \epsilon^{-1} \beta$.
Thus the amplitude of the longitudinal motion of the
substrate is $\mu b$ and the limiting case $\mu=0$ corresponds to
the transversal vibration, i.e. the results of
Ref.~\cite{PRE_vert} are reproduced. For the longitudinal
vibration it is necessary to set $|\mu|\gg 1$, $b\ll 1$, while keeping
the product, $B=\mu b$, finite in order to obtain the
corresponding formulas from
Secs.~\ref{sec:puls}-\ref{sec:dynamics}.

Representing all fields as the sums of the pulsatile and averaged parts according to
Eqs.~(\ref{rescaling})
we arrive at the following equations for
the pulsations:
\begin{subequations}
\label{puls_tilt}
\begin{eqnarray}
\tilde W_Z &=& -{\bf \nabla}\cdot \tilde {\bf U}, \ \Omega \tilde
{\bf U}_{\tau} = - {\bf \nabla} \tilde p + \tilde {\bf U}_{ZZ} ,
\\
\tilde p_Z &=&  -b\Omega^2 \cos \tau, \label{puls_p_z1}
\\
\tilde {\bf U} &=& b {\bf m} \Omega\sin\tau, \ \tilde W = 0 \ {\rm at} \ Z = 0, \\
\nonumber \Omega \tilde h_{\tau} &=& - \tilde {\bf U} \cdot {\bf
\nabla}\bar h + \tilde W, \\
 \tilde {\bf U}_Z &=&0, \ \tilde p = 0
\ {\rm at} \ Z = \bar h.
\end{eqnarray}
\end{subequations}
Here ${\bf m} \equiv \mu {\bf e}_x$. The averaged motion is
described by Eqs.~(\ref{aver_exp}).

Representing the solution of the boundary value problem
(\ref{puls_tilt}) in the form
\begin{subequations} \label{puls_compl_tilt}
\begin{eqnarray}
\tilde p &=& b \Omega {\rm Re} \left[q(X,Y,Z)\exp{\left(i\tau\right)}\right],\\
\tilde {\bf U} &=& b \Omega {\rm Re} \left[{\bf I}(X,Y,Z)\exp{\left(i\tau\right)}\right],\\
\tilde W &=& b \Omega {\rm Re} \left[K(X,Y,Z)\exp{\left(i\tau\right)}\right],\\
\tilde h &=& b {\rm Re} \left[H(X,Y)\exp{\left(i
\tau\right)}\right].
\end{eqnarray}
\end{subequations}
[cf. Eqs.~(20) in \cite{PRE_vert} and Eqs.~(\ref{puls_complex})]
and solving the equations for the amplitudes $q, {\bf I}, K$, and
$H$ we obtain (hereafter the bar over $\bar h$ is omitted again):
\begin{subequations}\label{sol_puls_tilt}
\begin{eqnarray}
\label{press_puls} q&=& \Omega\left(h-Z\right), \\
{\bf I}&=& i{\bf \nabla}h-i{\bf M} \frac{\cos\alpha (h-Z)}{\cos\alpha h}, \\
\nonumber K&=&-i\nabla^2
h\left[Z+\frac{\sin\alpha\left(h-Z\right)-\sin\alpha
h}{\alpha\cos\alpha h} \right] \\
&&+i{\bf \nabla} h \cdot
{\bf M} \frac{1-\cos\alpha Z}{\cos^2\alpha h},\\
H&=&-{\nabla} \cdot\left[ f\left(\alpha h \right)h{\bf \nabla}
h\right] + \frac{{\bf m}\cdot {\nabla} h}{\cos^2\alpha h},
\end{eqnarray}
\end{subequations}
where ${\bf M} \equiv {\bf \nabla} h + {\bf m}$ and
\begin{equation}
f(y)=1-y^{-1}\tan y.
\end{equation}
%
Due to the linearity of Eqs.~(\ref{puls_tilt}) the oscillatory
flow is a superposition of the motions generated by a vertical
[see Eqs.~(42) in \cite{PRE_vert}] and a horizontal
[Eqs.~(\ref{sol_puls})] vibration.

Referring to Sec.~\ref{ssec:puls-lim} as well as to Sec.~III of
Ref.~\cite{PRE_vert}, we do not present the limiting cases of the
solution (\ref{sol_puls_tilt}). Milestones of the stability
analysis for the pulsatile velocity are given in
Appendix~\ref{app:stab_tilt}.

Generally speaking the solution (\ref{sol_puls_tilt}) remains
valid even for the complex-valued $\mu$ and ${\bf m}$. This permits
to consider the vertical and horizontal vibration to be
out-of-phase, but leads to the cumbersome equation for the averaged
fields. Thus, hereafter we assume real ${\bf m}$.

\subsection{Amplitude equation and limiting cases} \label{sec:tilted_second}

\begin{figure}[!t]
\includegraphics[width=7.0cm]{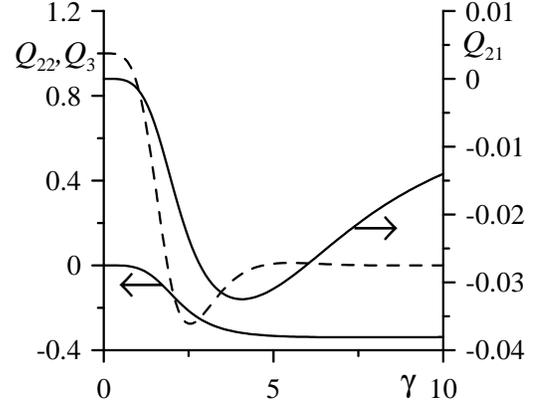}
\caption{Left axis -- the dependence of the coefficients $Q_{22}$
(solid line) and $Q_3$ (dashed line) in Eq.~(\ref{h-av_tilt}), on
$\gamma$. Right axis -- $Q_{21}\left(\gamma\right)$. }
\label{fig:Q23}
\end{figure}

Solving the averaged problem using the procedure similar to the one
in Appendix~\ref{app:aver_sol}, we arrive at the amplitude
equation for the averaged height:
\begin{subequations}\label{h-av_tilt}
\begin{eqnarray}
h_T&=&{\nabla}\cdot\left(\frac{1}{3}h^3 {\nabla} \Pi\right)-\frac{1}{2}b^2 \Omega^2 {\bf \nabla} \cdot {\bf Q}, \\
\Pi &=& \Pi_0+\frac{b^2\Omega^2}{2} {\rm Re}H,\\
\nonumber {\bf Q} &=&  h^3\left[Q_{21}\nabla^2 h {\bf M} + \left(
Q_{22}{\bf
M}+ \frac{1}{3} {\bf m}\right) \cdot {\bf \nabla \nabla} h \right] \\
&&+ h^2 Q_1 {\bf M} {\bf M} \cdot {\bf
\nabla} h,
\\
{\rm Re} H&=& -f_r h\nabla^2h-({\bf \nabla} h)^2 + Q_3{\bf
M}\cdot{\bf \nabla} h.  \label{ReH}
\end{eqnarray}
\end{subequations}
Here $\Pi_0$ and $Q_1$ are given by Eqs.~(\ref{Pi_0}) and
(\ref{Q1}), respectively,
\begin{eqnarray}
Q_{21}&=& 6q_1 -2q_2,\
Q_{22}=5q_1-q_2-\frac{1}{3},\\
Q_3&=& 2\frac{\cosh \gamma \cos\gamma+1}{\left(\cosh \gamma + \cos
\gamma\right)^2},\\
f_r&=&{\rm Re} f\left(\alpha
h\right)=1-\frac{\sinh\gamma+\sin\gamma}{\gamma\left(\cosh \gamma
+ \cos \gamma\right)}.\\
q_1 &=& \frac{\sinh
\gamma-\sin\gamma}{\gamma^3\left(\cosh\gamma+\cos\gamma\right)},\
q_2=\frac{\cosh\gamma-\cos \gamma}{\gamma^2\left(\cosh\gamma+\cos
\gamma\right)}.
\end{eqnarray}
The variation of the coefficients $Q_{21}, \, Q_{22}$, and
$Q_3$ with $\gamma$ is shown in Fig.~\ref{fig:Q23}, whereas $f_r$
can be found in Fig.~6(a) of Ref.~\cite{PRE_vert}.

For 2D case, when $h$ depends on $X$ and $T$ only, the amplitude
equation reduces to
\begin{subequations}\label{h-av_tilt2D}
\begin{eqnarray}
h_T&=&\left(\frac{1}{3}h^3\Pi_X\right)_X-\frac{1}{2}b^2 \Omega^2  Q_X, \\
\nonumber Q &=&Q_1\left(\gamma\right)h^2\left(h_X + \mu\right)^2
h_X \\
&&+ Q_2\left(\gamma\right) h^3 h_{XX} \left(h_X + \mu\right) +
\frac{1}{3}\mu h^3 h_{XX}
\end{eqnarray}
\end{subequations}
Here $Q_2\equiv Q_{21}+Q_{22}$, cf. Eq.~(31b) in
Ref.~\cite{PRE_vert}.

Equations (\ref{h-av_tilt}) coincide with Eq.~(\ref{h-av_t}) at
$|\mu|\gg 1, \, \mu b= B$ (the longitudinal vibration).
Equations~(\ref{h-av_tilt2D}) for $\mu=0$ (the vertical vibration)
coincide with Eqs.~(30) in Ref.~\cite{PRE_vert}, but the
corresponding 3D analogue reads
\begin{subequations}\label{h-av_vert}
\begin{eqnarray}
 h_T&=&{\nabla}\cdot\left(\frac{1}{3}h^3 {\nabla} \Pi\right)-\frac{1}{2}b^2 \Omega^2 {\bf \nabla} \cdot {\bf Q}, \\
\nonumber {\bf Q} &=&Q_1h^2\left({\bf \nabla} h \right)^2
{\bf \nabla} h \\
&&+ h^3\left[Q_{21}\nabla^2 h {\bf \nabla} h + Q_{22}{\nabla} h
\cdot {\nabla\nabla} h\right],
\end{eqnarray}
\end{subequations}
${\bf M}$ must be replaced with ${\bf \nabla h}$ in the definition of
$\Pi$, Eq.~(\ref{ReH}). It is clear that ${\bf Q}$ differs from
the one defined by Eq.~(43c) in Ref.~\cite{PRE_vert}. This
contradiction is caused by the calculation mistake in Ref.
\cite{PRE_vert}.
%
%

It is clearly seen that the amplitude equation (\ref{h-av_tilt})
contains several cross terms, which are linear with respect to
$\mu$ and proportional to the first derivative of $h$ with respect
to $X$. These terms remove the degeneracy with respect to the
replacement $X \to -X$. Only the invariance under the simultaneous
transformation $X \to -X$ and $\mu \to -\mu$ holds. Thus the
presence of the vertical vibration makes different the motion
along the positive and negative directions of the $X$ axis.
Indeed, the $X$-component of the oscillatory velocity is in-phase
with the vertical component, whereas the projection of the
pulsatile velocity on the $-X$ axis is in counter-phase. This
phase shift results in the difference after averaging.

The corresponding limits for the general case of the tilted
vibration give the following amplitude equations:

(i) low frequency  ($\Omega \ll 1$):
\begin{subequations}\label{h-av_tilt_low}
\begin{eqnarray}
\label{ampl_low_freq} h_T&=&\frac{1}{3}{\nabla}\cdot\left(h^3
{\nabla} \bar
\Pi\right)+b^2 \Omega^4 {\bf \nabla}\cdot{\bf Q}_l,\\
\nonumber {\bf Q}_l&=& \frac{h^7}{315}\left(\nabla^2 h {\bf M} +9
{\bf M}\cdot {\nabla
\nabla} h\right) \\
&&-\frac{h^6}{15}{\bf MM}\cdot {\bf \nabla} h,\\
\bar \Pi&=&\Pi_0-\frac{b^2\Omega^4}{15}{\bf
\nabla}\cdot\left(h^5{\bf M}\right).
\end{eqnarray}
\end{subequations}
For the vertical vibration this leads to
Eq.~(\ref{ampl_low_freq}) with
\begin{subequations}\label{h-av_vert_low}
\begin{eqnarray}
\nonumber {\bf Q}_l&=& \frac{h^7}{315} \left(\nabla^2h {\bf
\nabla}h+9\nabla
h\cdot {\bf \nabla\nabla} h\right) \\
&&-\frac{h^6}{15}
\left(\nabla h\right)^2 {\bf \nabla}h,\\
\bar \Pi&=&\Pi_0 - \frac{b^2\Omega^4}{15}{\bf \nabla}\cdot (h^5
{\bf \nabla}h)
\end{eqnarray}
\end{subequations}
instead of Eq.~(44) in \cite{PRE_vert}. For the horizontal
vibration Eq.~(\ref{h-av_low}) is reproduced.

(ii) high frequency ($\Omega \gg 1$):
\begin{subequations}\label{h-av_tilt_high}
\begin{eqnarray}
h_T&=&\frac{1}{3}{\nabla}\cdot\left\{h^3 {\nabla}
\left[\Pi+\frac{b^2 \Omega^2}{4} \left({\bf
\nabla}h\right)^2\right]\right\},\\
\Pi&=&\Pi_0-\frac{b^2\Omega^2}{2}{\bf \nabla}\cdot\left(h{\bf
\nabla}h\right).
\end{eqnarray}
\end{subequations}
It is obvious that the longitudinal component of the vibration has no
impact in this limiting case (see the explanation in
Sec.~\ref{ssec:limits}). Therefore, the obtained expression
coincides with the corresponding equations obtained in
Ref.~\cite{Lapuerta} as well as Eqs.~(46) and (47) in
\cite{PRE_vert}.

\subsection{Stability analysis of the flat surface}

Representing $h$ in the form $1+\xi$ and linearizing
Eqs.~(\ref{h-av_tilt}) with respect to a small perturbation $\xi$
results in
\begin{eqnarray}
\nonumber \xi_T&=&\frac{1}{3}\nabla^2
\left[\left(G_0-3A\right)\xi-\left(C+\frac{b^2\Omega^2}{2}f_r\right)\nabla^2
\xi\right]\\
&&-\frac{b^2\Omega^2}{2} \left(\frac{3Q_2+1-Q_3}{3}\mu \nabla^2
\xi_X+\mu^2 Q_1 \xi_{XX}\right).
\end{eqnarray}
Here $\gamma$ must be replaced by $\gamma_0$ in $Q_j, \ j=1,2,3$
and $f_r$, because all the coefficients are calculated for the
unperturbed state, $h_0=1$. Substituting $\xi$ proportional to
$\exp\left(ik_X X +ik_Y Y-\lambda T\right)$ and separating real
and imaginary parts of the decay rate we arrive at
\begin{eqnarray}
\nonumber \lambda_r&=&\frac{1}{3}k^2\left[G_0-3A+k^2\left(C+\frac{b^2\Omega^2}{2}f_r\right)\right]\\
&&-\frac{b^2\Omega^2}{2}\mu^2
Q_1 k_X^2,\\
\lambda_i&=&-\frac{b^2\Omega^2}{2}\mu k^2
k_X\left[\frac{\sinh^2\gamma-\sin^2\gamma}{3\left(\cosh\gamma+\cos
\gamma\right)}+Q_2\right].
\end{eqnarray}

It can be readily seen that the impact of tilted vibration on the
real part of the decay rate, $\lambda_r$, is the superposition of impacts from
the vertical and horizontal vibration
[cf. Eq.~(72) in \cite{PRE_vert} for the former case and
Eq.~(\ref{decay}) for the latter one]. The additional terms,
linear with respect to $\mu$, provide only the imaginary part of
$\lambda$.

Thus the vertical component of the vibration is unimportant for the
longwave perturbations (with $k\rightarrow 0$), and in this case the tilted vibration,
similar to the horizontal one, decreases the stability
threshold (unless $Q_1<0$, see the discussion of Eq.~(\ref{decay})).

Again, the 2D perturbations ($k_Y=0, \, k_X=k$) are critical for
$Q_1>0$ and longitudinal rolls ($k_X=0$) are critical for narrow
intervals of $\Omega$, where $Q_1<0$. However, the latter case
seems unrealistic as very high frequencies are needed.

In confined systems with a discrete spectrum of $k$ the
competition of the stabilizing effect of the vertical vibration and
the destabilizing effect of the horizontal one takes place.

It should be noted also that the emergence of the imaginary
part of $\lambda$ is the indicator of the averaged transport in
the system. Indeed, the perturbations are stationary in a
reference frame moving along the $X$ axis with a constant velocity
$\lambda_i/k_X$. However, such a longitudinal drag is not limited
by the transport of perturbations: any admixture can be spread
over the system by means of the tilted vibration. Thus the tilted
vibration seems to be the novel way to transport microparticles
or molecules, which is important in many microfluidic
applications.

\section{Summary}\label{sec:summary}
We consider a thin liquid film on a planar horizontal substrate subjected to a high
frequency vibration. In the absence of a vibration, the van der
Waals attraction to the substrate destabilizes the film and causes
its dewetting. In contrast to conventional averaging method, we
assume that the period of the vibration is comparable to the time
of viscous relaxation of perturbations across the layer. This
allows us to apply the averaging method to the ultra-thin films.
Such analysis was first developed in Ref.~\cite{PRE_vert}, where
the vertical vibration is considered and is shown to enhance film
stability.

This work is a natural extension of Ref.~\cite{PRE_vert}.
We consider, separately, the longitudinal and the tilted vibration. In the former
case the {\em finite amplitude} of the vibration results in
destabilization of the layer. There is also a sequence of narrow
intervals of the vibration frequency, where stabilization occurs
in the two-dimensional problem. However, the frequency must be very
high (at least $300 \, {\rm MHz}$ for a water layer of the
thickness $1000 \ \AA$).

Another effect of the longitudinal vibration is the emergence of the
supercritical branching at the sufficiently high intensity of the vibration. In
this case the deformed free surface becomes stable, i.e. the
instability of the flat surface does not necessarily lead to a
rupture.

For the tilted vibration the longitudinal (destabilizing)
component of the pulsatile velocity is shown to be dominant. The
only case, where the competition of the vertical and the horizontal
vibration occurs, is the almost vertical vibration of large
amplitude. For this case the averaging procedure is carried out
and the corresponding amplitude equation is obtained. This
analysis allows to correct the three-dimensional generalization of the amplitude
equation for the vertical vibration \cite{PRE_vert}.

Linear stability analysis in the framework of the amplitude
equation indicates that both destabilization and stabilization of
the flat surface are possible in the case of tilted vibration.
Stabilization takes place only in the confined systems, when the
spectrum of perturbations is discrete and bounded from below.

Besides, the small perturbations are oscillatory, i.e. the drag
takes place for the tilted vibration. This property can be very
important for many microfluidic applications since some admixtures
can be transported in the same manner as the perturbations.

\section{Acknowledgements}
S.S. is partially supported by the Foundation ``Perm
Hydrodynamics''.

\appendix
\section{Solution of the averaged problem. Longitudinal
vibration}\label{app:aver_sol}

The solution of the averaged boundary value problem,
Eqs.~(\ref{aver}), is sought in the form:
\begin{eqnarray}
p&=&-G_0Z+\Pi_0,\\
\label{U_av} {\bf U}&=&{\bf U}_c+\frac{b^2\Omega^2}{2}U_v{\bf
e}_x,
\end{eqnarray}
where $\Pi_0$ is given by Eq.~(\ref{Pi_0}),
$$
{\bf U}_c=\frac{1}{2} Z\left(Z-2h\right) {\bf \nabla} \Pi
$$
is the ``conventional" part of the longitudinal velocity, and its
``vibrational'' part is determined by the boundary value problem
\begin{subequations} \label{Uv}
\begin{eqnarray}
U_v^{\prime\prime}&=&{\rm Re}\left(II_X^*+KI_Z^*\right),\\
U_v&=&0 \ {\rm at} \ Z=0, \\
U_v^\prime&=&-\frac{1}{\Omega}{\rm Re}\left(HI_{ZZ}^*\right) \
{\rm at} \ Z=h.
\end{eqnarray}
\end{subequations}
It can be readily shown that
\begin{equation}
U_v={\rm Re}\int_0^Z KI^* {\rm d}Z- \int_0^Z {\rm d} \xi
\partial_X \int_\xi^h |I|^2 {\rm d}\eta
\end{equation}
is the solution of Eqs.~(\ref{Uv}).

Substituting Eq.~(\ref{U_av}) with both known parts, ${\bf U}_c$
and $U_v$, into Eq.~(\ref{int_aver}) one can obtain
Eq.~(\ref{h-av_t}) after straightforward (but cumbersome) integration.

\section{Stability of the pulsatile flow. Tilted vibration}\label{app:stab_tilt}

Here we only briefly analyze the stability of the pulsatile flow
(\ref{sol_puls_tilt}) as this analysis is quite similar to one
given in Sec.~VII of Ref.~\cite{PRE_vert} and
Sec.~\ref{sec:stab_puls} of the present paper.

In order to study the behavior of short-wavelength
perturbations, we again return to the unscaled system of
equations~(\ref{base_eq}) and (\ref{base_bcs}) with the
above-mentioned changes: (i) and (ii), see
Sec.~\ref{sec:tilted_first}. The perturbed fields are
\begin{equation}\label{pert_puls_tilt}
{\bf v}=\tilde {\bf U} + \epsilon \tilde W {\bf e_z}+{\bf V}, \
p=\epsilon^{-1}\tilde p+\bar p+P, \ h=\bar h +\epsilon \tilde h +
\Upsilon.
\end{equation}
The only difference of Eqs. ~(\ref{pert_puls_tilt}) from Eqs.~(\ref{pert_puls}) is the presence of the
oscillatory part of the pressure, according to
Eq.~(\ref{press_puls}). In Eq.~(\ref{pert_puls_tilt}) the
oscillatory parts are given by Eqs.~(\ref{sol_puls_tilt}).
Below we again omit the bar over $\bar h$.

Repeating the same procedure of the method of frozen coefficients,
we arrive at the following set of equations governing the
small-amplitude perturbations:
\begin{subequations}
\label{pert_puls_eq_tilt}
\begin{eqnarray}
W^{\prime}&=&-{\bf \nabla} \cdot {\bf U},
\\
{\bf U}_t&=&- {\bf U}_0 \cdot {\bf \nabla} {\bf U} - W {\bf U}_0^{\prime}   - {\bf \nabla}P +
\nabla ^2 {\bf U}+{\bf U}^{\prime\prime}, \\
W_t &=&- {\bf U}_0\cdot {\bf \nabla} W  - P^{\prime} + \nabla ^2 W +W^{\prime\prime},\\
{\bf U} &=& W=0 \ {\rm at} \ z = 0, \\
\nonumber \Upsilon_t &=&- {\bf U}_0 \cdot {\bf \nabla}\Upsilon + W,\  {\bf U}^{\prime}+\nabla W=-{\bf U}_0^{\prime\prime} \Upsilon,\\
\label{Ppet}\nonumber P &=& \left(\frac{b\Omega^2}{\epsilon} \cos\Omega t + G+
\frac{3A}{h^4}\right)
\Upsilon  - \frac{C}{\epsilon^2} \nabla^2 \Upsilon\\
\label{dyn_puls_tiltt}&&+2\left(W_z-{\bf U}_0^\prime \cdot \nabla
\Upsilon\right) \ {\rm at} \ z = h.
\end{eqnarray}
\end{subequations}
Here the primes denote the $z$-derivatives and the base oscillatory velocity is
\begin{equation}\label{2Dbase}
{\bf U}_0=b\Omega{\rm Re} \left[{\bf I}(z)\exp\left(i\Omega
t\right)\right].
\end{equation}
This boundary value problem is a straightforward 3D extension of
Eqs.~(\ref{pert_puls_eq}), with two exceptions. Firstly, there is an additional term
at the right-hand side  of
Eq.~(\ref{Ppet}), which comes from $-\tilde p_z \Upsilon$.
Secondly, the base oscillatory velocity is 2D.

As it is shown in Ref.~\cite{PRE_vert} [see Eqs.~(52) there], the
presence of the two asymptotically large terms in
Eq.~(\ref{Ppet}) allows to split the stability problem
into two different problems: (i) the Faraday instability and (ii)
the instability of the oscillatory flow with the undeformable free
surface.

(i) The solution is based on the results by
Mancebo and Vega \cite{Mancebo}. This analysis gives Eq.~(53) of
Ref.~\cite{PRE_vert}:
\begin{equation}
b_c=\frac{\sqrt{C}\Phi_{MV}\left(\Omega
h^2\right)}{\Omega^2h^{5/2}},
\end{equation}
where $\Phi_{MV}$ is the function given in Fig.~4 of
Ref.~\cite{Mancebo}.

(ii) It is easy to see that the stability problem for 2D base flow
(\ref{2Dbase}) can be reduced to Eqs.~(\ref{Orr_Zommer_resc}).
Indeed, after the separation of the longitudinal coordinates according
to $f(x,y,z,t)=\hat f(z,t)\exp(i{\bf k}\cdot{\bf R})$ and the
obvious transformation
$$
\hat f \rightarrow \hat f \exp\left[-i b \cos \Omega t{\bf k}\cdot
\nabla h\right],
$$
the base velocity (\ref{2Dbase}) becomes locally 1D:
\begin{equation}\label{1Dbase}
{\bf U}_0=-b\Omega{\bf M}{\rm Re} \left[i \frac{\cos\alpha
(h-Z)}{\cos\alpha h} e^{i\Omega t}\right].
\end{equation}
Here $f$ is each of the fields $\{P, W, {\bf U}=(U,V),
\Upsilon\}$, and $\bf k$ and $\bf R$ are 2D vectors in plane
$x-y$.

Thus, we again  exclude the
constant part of the velocity, i.e. the first term in $\bf I$ by the
periodical in time transformation similar to
Eq.~(\ref{transform}) (see Sec.~\ref{ssec:Pois_stab}). Now one can redirect the local $x$ axis
along the vector $\bf M$ and introduce the streamfunction in order
to obtain exactly the same Orr-Sommerfeld problem as
Eq.~(\ref{Orr_Zommer}). It has been shown in
Sec.~\ref{ssec:Pois_stab} that the latter flow is stable.


\begin{thebibliography}{99}

\bibitem{SHJ}
R.~Seemann, S.~Herminghaus, and K.~Jacobs, J. Phys.: Condensed
Matter {\bf 13}, 4925 (2001).

\bibitem{Oron}
A.~Oron, S.~H.~Davis, and S.~G.~Bankoff, Rev. Mod. Phys. {\bf 69},
(1997) 931.

\bibitem{Eggers}
J.~Eggers,
Rev. Mod. Phys. {\bf 69}, (1997) 865.




\bibitem{AverBook}
J.~A.~Sanders and F.~Verhulst, \textit{Averaging methods in
nonlinear dynamical systems} (Springer-Verlag, New York, 1985).

\bibitem{TVC}
G.~Z.~Gershuni and D.~V.~Lyubimov, \textit {Thermal Vibrational
Convection} (Wiley, New York, 1998).

\bibitem{LChbook}
D.~V.~Lyubimov, T.~P.~Lyubimova, and A.~A.~Cherepanov,
\textit{Dynamics of interfaces in vibration fields} (Fizmatlit,
Moscow, 2004) (in Russian).

\bibitem{wolf-70}
G.~H.~Wolf, Phys. Rev. Lett. {\bf 24}, 444 (1970).

\bibitem{Mancebo}
F.~J.~Mancebo and J.~M.~Vega, J. Fluid Mech. {\bf 467}, 307
(2002).

\bibitem{Hocking_Davis}
L.~M.~Hocking, S.~H.~Davis, J. Fluid Mech. {\bf 467}, 1 (2002).

\bibitem{Oron_Gottlieb}
A.~Oron, O.~Gottlieb, Phys. Fluids {\bf 14}, 2622 (2002).

\bibitem{Lapuerta}
V.~Lapuerta, F.~J.~Mancebo, and J.~M.~Vega, Phys. Rev. E {\bf 64},
016318.

\bibitem{Thiele}
U.~Thiele, J.~M.~Vega, and E.~Knobloch, J. Fluid Mech. {\bf 546},
61 (2006).


\bibitem{PRE_vert}
S.~Shklyaev, M.~Khenner, and A.~A.~Alabuzhev, Phys. Rev. E {\bf
77}, 036320 (2008).

\bibitem{squire-33}
H.~B.~Squire, Proc. R. Soc. London, Ser. A {\bf 142}, 621 (1933).


\bibitem{Singer}
B.~A.~Singer, J.~H.~Ferziger, and H.~L.~Reed, J. Fluid Mech. {\bf
208}, 45 (1989).

\bibitem{Straatman}
A.~G.~Straatman, R.~E.~Khayat, E.~Haj-Qasem, and D.~A.~Steinman,
Phys. Fluids {\bf 14}, 1938 (2002).




\bibitem{Davis_annu}
S.~H.~Davis, Annu. Rev. Fluid Mech. {\bf 7}, 57 (1976).

\bibitem{Kerczek}
C.~von Kerczek and S.~H.~Davis, J. Fluid Mech. {\bf 62}, 753
(1974).


\end{thebibliography}
\end{document}